\documentclass[a4,12pt]{article}
\usepackage{graphics}
\usepackage{epsfig}
\input epsf

\begin{document}

\centerline{\bf \huge{Adjustment and social choice }}

 \bigskip
\centerline{\bf G{\'e}rard~Weisbuch and Dietrich Stauffer$^2$} 

\centerline{Laboratoire de Physique Statistique$^4$}

\centerline{de l'Ecole Normale Sup{\'e}rieure, }
 
\centerline  {24 rue Lhomond, F 75231 Paris Cedex 5, France. } 

\bigskip
{\small $^2$Institute for Theoretical Physics, Cologne University, D-50923 K\"oln, 
Germany}

\noindent
\bigskip
{\small $^4$Laboratoire associ{\'e} au CNRS (URA 1306), {\`a} l'ENS 
et aux Universit{\'e}s Paris 6/7 }

\bigskip
\centerline{e-mail: weisbuch@lps.ens.fr}

\begin{abstract}
  We discuss the influence 
of information contagion on the dynamics of choices
in social networks of heterogeneous buyers.
Starting from an inhomogeneous cellular automata model
of buyers dynamics, we show that when agents try to
adjust their reservation price, the tatonement process does not 
converge to equilibrium at some intermediate market share
and that large amplitude fluctuations are actually observed.
When the tatonnement dynamics is slow with respect
to the contagion dynamics, large periodic oscillations
reminiscent of business cycles appear.
\end{abstract}

\section{Introduction}
\label{sec:intro}

 Bubbles in financial market are one of the
most spectacular stylised fact in contradiction
with General Equilibrium Theory. Economists
and ``econophysicists'' have also noticed that the spectral properties
of stock, commodities and foreign exchange return series   
were far from Gaussian noise: return series display 
scale invariance, a property that physicists and economists 
relate to cooperativity (interactions) among the agents.
 To our knowledge there have been few explicit 
models of the phenomenon and our aim in this contribution
is to discuss one of the simplest theoretical approaches.

 Basically our model couples interaction among agents
with their tatonnement procedure to get ``fair'' prices.
Both processes have attracted the attention of modelers,
see e.g. F\"olmer (1974), Galam (1982, 1991) and Orlean (1995)
 for interactions and herding behaviour
and Lesourne (1992) for adjustment.
\begin{itemize}
\item We suppose that agents are not independent and that their
individual choices are influenced by the choice of their neighbours
for whatever reason, externalities or information;
\item Agents initially have some reservation price
(the highest price they would accept for the good) 
that they adjust to match offers.
\end{itemize}

  In a previous attempt to understand these phenomena,
we used a percolation
model to describe the information contagion dynamics
(Solomon et al 2000, Goldenberg et al 2000, Weisbuch and Solomon
2000, Weisbuch et al 2001). Agents are nodes of a lattice and are 
susceptible to purchase a good when its quality meets their
expectations after one of their neighbours purchased.
 Adjustment of qualities and expectations are 
based on a standard tatonnement process which we describe further in
section 2.2. The resulting dynamics were checked by the observation
of time series of purchase, patterns of purchase, and Fourier
transform of the time series. A distinctive feature of cooperativity
was the observed $1/f^2$ spectrum of the purchase time series.
 Comparable results were obtained by Plourabou\'e etal (1998)
and Steyer and Zimmermann (2000) who reported $1/f$
noise when the slow dynamics is Hebbian learning.

  One of the purposes of the present study is to check the genericity
of the previously modeled behavior: we only have a very indirect
knowledge of the reasons why agents decide to purchase a good or not 
to purchase it, not to mention the specific algorithm they would use
to survey their neighbours before taking a decision. The 
previously used percolation
approach supposes that the purchase by one of their neighbours is
sufficient to provide a full knowledge of the quality of the product,
while the ``counter'' (or voting) dynamics presented here implies that agents
survey all their neighbours to take some average opinion.
There might be specific situations for which one
or the other scheme would make more sense, but
in general we would like to know
how different would the resulting global dynamics be.   

 The paper first describes the model. We then give
simulation results for the two different dynamical regimes.
 After some study of the influence of parameters 
we discuss in the conclusion the relevance of the 
observed dynamics to business cycles and financial markets.

\section{The INCA model}
\subsection{Information contagion}
 A rather standard model of information contagion
is based based on cellular ``counters''
(also called voters dynamics).
  Let us consider cellular automata: binary agents occupy sites of a
two dimensional lattice. The decision rule, buy or not buy,
corresponding to state $S_i=1$ or to state $S_i=-1$,\footnote{the
  choice 1 or 0 for buy or not buy, is more standard in
economics, but our choice, inspired from physics, respects 
symmetry and thus makes computations and mathematical expressions simpler.}  
is based on some combination of private information
and information coming from neighboring (on the lattice) agents.
The private information of agent $i$ is some threshold
$\theta_i$, and information incoming from neighbours is 
simply the sum of their individual states. At each time step
one agent randomly selected updates its state by comparing
the sum of its neighbours' states to its threshold:
\begin{equation}
  S_i =1 \quad \mathit{iff} \quad \sum_j S_j > \theta_i
\end{equation}
Otherwise $S_i = -1$.

   Homogeneous counters automata, with identical thresholds, are simple 
cellular automata which dynamical attractors depend on the amplitude 
of the threshold as compared to the number of neighbours. Lower  
thresholds (less than -2) give homogeneous attractors with state
1 for all automata,  higher  
thresholds (more than +2) give homogeneous attractors with state
-1 for all automata. Intermediate threshold values give coexisting domains
of plus and minus ones which size depends upon
 thresholds and initial conditions (see e.g. Weisbuch 1991 for more details).

  But we are interested here in inhomogeneous counters
with different thresholds: INCA\footnote{INhomogeneous Cellular
  Automata (Vichniac 1986).}
 are disordered systems and their dynamical properties 
reflect their disordered structure: one
observes stable dynamical regimes 
separated by finite width phase transitions for which attraction
basins depend up on the particular realisation of disorder.
 Weisbuch and Boudjema (1999) have shown for instance that even under
a favourable average threshold which would predict
invasion by +1's in a sea of -1's, the detailed positions
of the automata initially at state +1 are important.
The phenomenon is well-known in the 
physics of phase transitions: germs are initially necessary to
the growth of a stable phase in conditions of 
supercooling for instance.  
 
\subsection{The adjustment model}
  
  Let us now be more specific about the interpretation
of the threshold in terms of economics. The threshold which
drives the buying behavior of the agent can be seen as 
a price difference between how much a seller would like to get 
from a product $p_s$ and how much $p_b$ 
a buyer is ready to pay for it
when the signal from neighbours would cancel
(in other words with an equal number of buyers
and non-buyers in the agent neighborhood).
\begin{equation}
  \theta_i = p_s - p_b .
\end{equation}
A positive threshold would prevent purchase, a negative
one would allow it. The presence of other purchasers in the
neighborhood would favour purchase. 
(Of course prices have to be expressed in some units
consistent with equation 1. Two units in price difference
corresponds to a threshold change of one extra neighbour).

 The adjustment process now can be simply stated as:
 \begin{itemize}
 \item  When an agent did not purchase upon update
($S_i=-1$), she decreases her threshold with the hope to 
be able to be in a better condition to purchase next time;
  \item in the opposite case she increases it.
 \end{itemize}
  The algorithm is above described as adjustment of a buyer
reservation price,
but a symmetrical reasoning for a seller would give the same
threshold dynamics. In fact we cannot directly suppose 
a symmetrical reasoning on the buyer side: it makes a difference
whether there is only one seller (monopoly)
and how fast he would react, or 
whether we have several buyers and what are the trading relationships
between buyers and sellers. Let us then supppose for
 the sake of simplicity that only buyers adjust their 
reservation price.
 
 The adjustment dynamics in the absence of any coupling
between agents would be similar to the simple mechanism
 describe e.g. by Laffond and Lesourne (1992) and would yield a similar 
clearing of the market as described in Lesourne's book (1992).
The difference here is that we are interested in 
multiple purchases by agents and that we couple
adjustment and contagion dynamics.

 \section{Simulation results}  

\subsection{The slow adjustment regime}
  Before a full study of parameters and variants
let us try to understand the simplest case. An
obvious parameter of the model is how fast the threshold is adjusted
with respect to the buying propagation dynamics.
Let us suppose that at every update, the adjustment amplitude is
random and uniformly distributed on $[0,q]$.
We further refer to $q$ as the adjustment rate.
Its magnitude has to be compared with the number of neighbours
taken into account in the simulation: we used four.
For slow adjustment rate such as $q=0.1$, we expect the dynamics of adjustment
to be slow with respect to the diffusion dynamics.
Let us define the relative adjustment rate $\rho$ as the 
ratio between the average adjustment, here $q/2$, and the
difference between thresholds such that isolated individuals
among a neighbourhood of either buyers or non-buyers
 would take an opposite view to their neighbours, here 8.
 The slow adjustment regime is such that:

 \begin{equation}
   \rho < \frac{2}{L}
 \end{equation}
where $L$ is the width of the square lattice. The rhs term is
the inverse of the time necessary to propagate a position,
buying or not buying, across the net, 
under the most favourable threshold condition 
 (the term 2 is due to the fact that we use
periodic boundary conditions). 

  The simulation conditions are then:
  \begin{itemize}
  \item A square lattice of dimension $L^2$ (e.g. $20 \times 20$ for figure 1);
  \item random\footnote{at each time step one node randomly
chosen is updated according to equation 1 for its state and 
section 2.2 for its threshold} updating based on the described algorithms;
 \item The initial configuration of agents is random
for the binary state and thresholds are uniformly distributed
on [-1,1]. 
  \end{itemize}

  Figure 1 is a time plot of the average state of agents
(fraction of buyers)
and average threshold.

 \begin{figure}
\centerline{
\epsfxsize=80mm
\epsfig{file=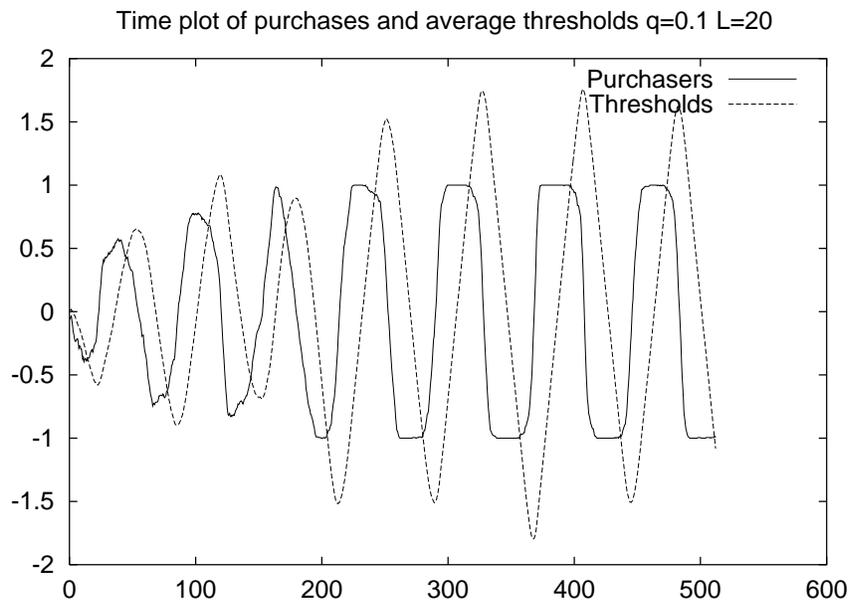,angle=270,width=120mm}}
\caption{
Time evolution of the average state of agents and average threshold,
in the slow adjustment regime.
(average state=1, everyone buys, average state=-1, no-one buys.).
Unit time correspond to updating each site once on average).}
\end{figure}

 The regular oscillations of agents states and
average thresholds obtained at long time give some indication 
of the processes which control the dynamics. 
Note that both quantities display relaxation 
rather than sinusoidal oscillations, threshold
varying as a triangular  wave and purchases
more like a square wave which saturates at maximum
(all or no agents purchasing). The two quantities 
have a phase shift of $\pi/2$: extreme variations
of the average threshold occur when purchases saturate.
These observations plus
direct online observations of the lattice dynamics 
for purchase and threshold can be easily understood.

 Once the lattice is in a saturated condition,
say everyone buying, an isolated agent who would
choose not to buy needs a threshold much higher   
than if she were surrounded by non-buyers. The system
has to ``wait'' until thresholds which were low during
the rise of the purchasing behaviour rise again
to allow the apparition of isolated non-buyers. Hence
the straight part of the average threshold evolution
corresponding to its slow and regular increase.
But as
soon as isolated non-buyers are present, their neighbours
need a lower threshold to switch to 
no-purchase; a wave of no-purchase propagates
across the lattice.
Hence the fast switch observed on the purchase
time evolution: online observation display the 
growth of non-purchaser germs surrounded by disappearing
domains of purchasers. One single sweep from
purchasing to non-purchasing is the equivalent
of a phase transition for which germs are needed
(first order phase transitions).
 The phenomenon is symmetrical for purchase and no purchase, 
hence the observed oscillations.

\subsection{The fast adjustment regime}

 With large networks and fast adjustment rates, the global
synchrony between all agents on the lattice is destroyed.
Agent states (purchase/no-purchase) and thresholds
display small homogeneous domains on the lattice.
Because of the randomness of the updating process,
some agents easily reach ``eccentric'' 
(opposed to their neighbours) positions and 
many domains out of phase start growing in different
places\footnote{By contrast,
 in the slow adjustment regime, 
many time steps are needed to 
sweep from the lower to the upper threshold, and
the standard deviation of the threshold is reduced
by the summing process, thus allowing synchrony of
agents behaviour.}.
This behaviour is observed with online simulations
and displayed on figure 2 at time 100 000.

  The change of dynamical regime with adjustment rates $q$ 
is very smooth and corresponds to a crossover
rather than to a phase transition.

 \begin{figure}
\centerline{
\epsfxsize=50mm
\epsfig{file=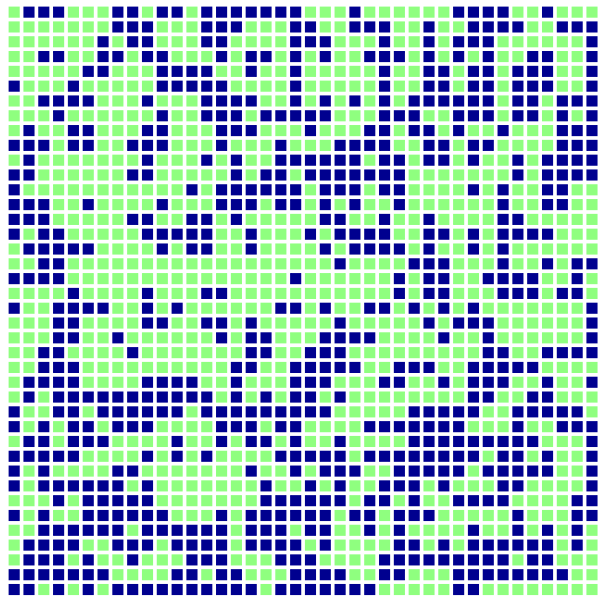,angle=270,width=80mm}}
\caption{
Pattern of behaviour at time 100
 in the fast adjustment regime.  Adjustment rate
$q$ is 0.7. Grey squares correspond to
buyers, black squares to non-buyers. (In the slow 
adjustment regime domains would be much larger. Sometimes,
consensus across the lattice is achieved).}
\end{figure}

Because of domains asynchrony, oscillations are relatively smaller and less 
regular in amplitude in the fast adjustment regime
 than in the slow adjustment regime
as observed in figure 3.
Saturation of the (no)purchaser fraction is never reached. 

 \begin{figure}
\centerline{
\epsfxsize=80mm
\epsfig{file=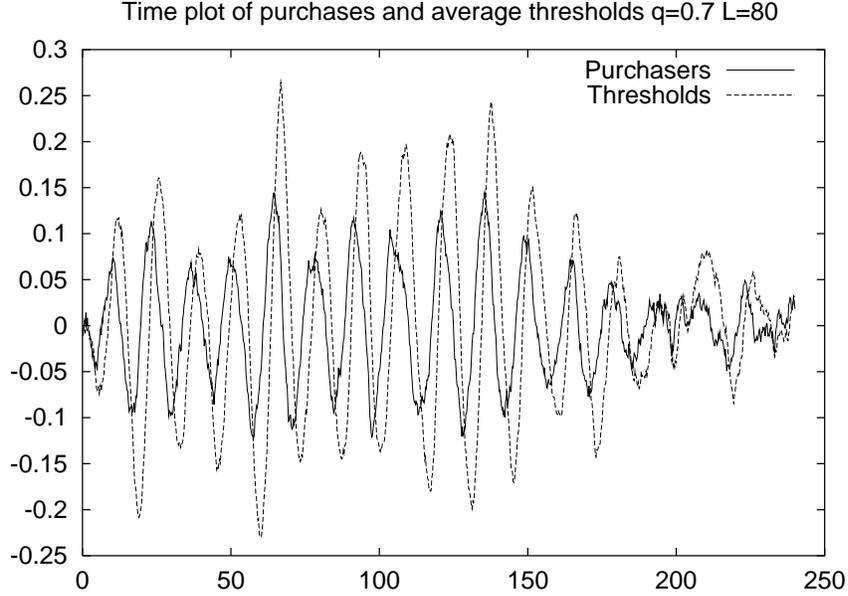,angle=270,width=120mm}}
\caption{
Time evolution of the average state of agents
and average threshold
 in the fast adjustment regime. Oscillation are smaller and less 
regular in amplitude than in the slow adjustment regime.}
\end{figure}

Another way to monitor inhomogeneity is
to check the spatial autocorrelation
function of states.
\begin{eqnarray}
  C(d)=\frac{\sum_{i}S(i)S(i+d)-(\sum_{i}S(i))^2}{L^2}
\end{eqnarray}
where $i$ is the position of lattice sites and $d$ 
the translation distance.
 $L^2$ is the number of agents.

 Figure 4 displays the autocorrelation function for 
different lattice sizes in the fast adjustment regime
when the adjustment rate $q=0.7$.
Its fast decay, over some 4 lattice sites,
tells us that the average linear dimension of purchasing
or no-purchasing domains is of order 4 to be compared
with the patterns observed on figure 2.
A striking result is that although lattice linear sizes
change by a factor 8, the autocorrelation function
are very similar, implying that the average size
of domains is independent of lattice size in the fast adjustment
regime.
In other words, figure 2 is statistically representative
of any part of a larger lattice.

 \begin{figure}
\centerline{
\epsfxsize=80mm
\epsfig{file=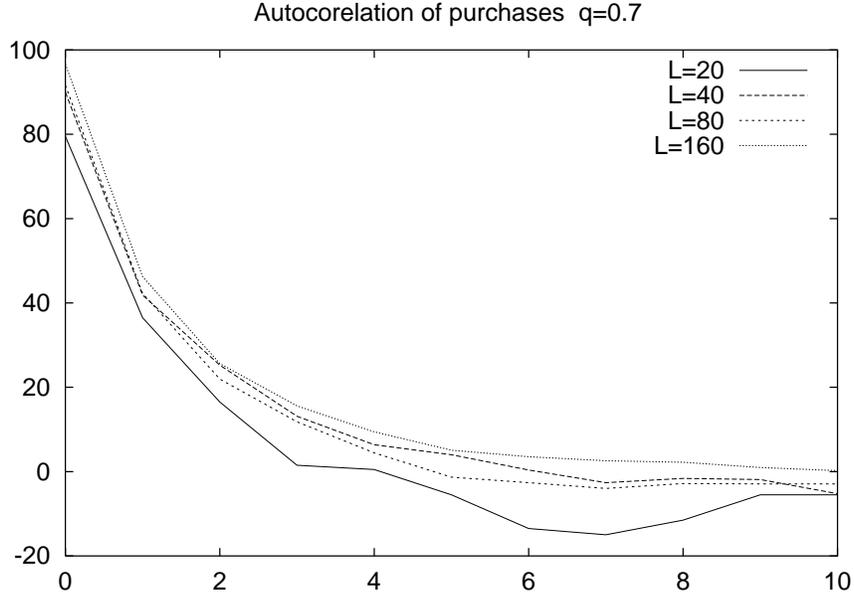,width=120mm}}
\caption{
Spatial autocorrelation
function of the average fraction of buyers
for lattice sizes varying from $20 \times 20$
to $160 \times 160$. Abscissa is distance in units of lattice
spacing. Note that the correlation length,
given by the distance at which the autocorrelation
is zero is nearly invariant with lattice size.}
\end{figure}

{\bf Scaling}

  The system has only two parameters, $L$ the 
lattice dimension and the average adjustment rate $\rho$.
We would like to know how the characteristic variable
of the dynamics, frequency and amplitude of the oscillations,
and their space dependence through the autocorrelation function,
vary with $L$ and $\rho$.

 Frequencies are surprisingly stable over time
and from sample to sample as opposed to magnitudes.
A direct measure on time plots of oscillations shows that 
periods $T$ vary as:
\begin{eqnarray}
  T \simeq \frac{10 L^2}{q}
\end{eqnarray}
 This result has a very simple interpretation.
The factor $L^2$ is simply the number of agents.
 The period is close to the number of agents  
multiplied by a time which scales with
average time it takes for threshold to
switch between extremal values of -4 and 4.
 The threshold dynamics is the rate limiting
step of the overall dynamics. 

 As seen in figure 3, amplitudes display a lot of variations.
 A simple way to average
them on time is to measure power, namely the time averaged squared
amplitudes. Even with time averaging over some 800 periods,
power values had to be further averaged over several runs (9 in our 
measurements)
to further reduce noise. A first result is that for larger values
 $\rho$  values, $\rho>5$, average power scales as  $L^2=N$ 
the number of agents.
If agents behaviour were oscillating in phase, we would expect
power to scale in $N^2$. The scaling in $N$ implies 
that $N/s$ patches of constant size $s$ oscillate independently
giving:
 \begin{eqnarray}
  P \sim \frac{N}{s} P_s \sim N s \sim \frac{N}{q^2}
\end{eqnarray}   
where $P_s$ is the power of one patch, proportional to $s^2$.
This interpretation is consistent with our interpretation
of autocorrelation measurements and the observation of
small domains. The scaling of $s$ in $q^{-2}$
is obtained from the equivalence between the time
it takes for the social influence to sweep the patch 
and the time it takes to the threshold
adjustment to sweep between the extreme values.
 
Figure 5 displays the rescaled inverse power (i.e.  $\frac{P}{N}$) as
a function of $q$ the maximum adjustment rate ($q=8\rho$)
for $N$ varying from 400 to 6400. The collapse of the 
three curves above $q=0.6$ is good, the quadratic scaling 
in $q$ is approximate.

 \begin{figure}
\centerline{
\epsfxsize=80mm
\epsfig{file=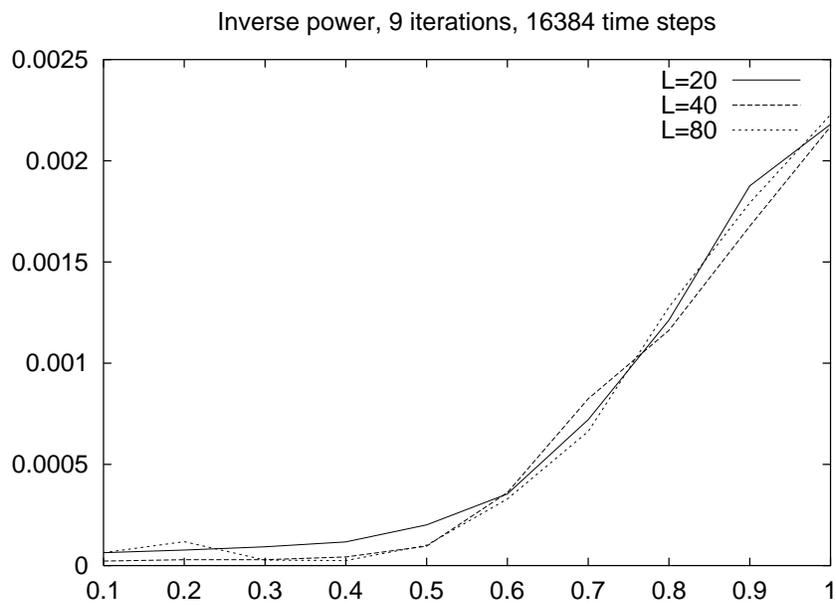,width=120mm}}
\caption{ Rescaled inverse power in the fast adjustment regime,
for several network sizes, $L=20,40,80$ as a function
of the adjustment rate $q$. When $q>0.5$
one observes a good collapse of simulation data 
for the rescaling in $N$
and a quadratic variation in $q$.  
}\end{figure}

Figure 6 displays the Fourier power spectrum
 of the time series of agent states when $q=1$. The
large peak around abscissa 30 corresponds 
to a frequency of 10 iterations per agent.
At larger frequencies, the long tail
corresponds to a $1/f^2$ noise. Small scale 
correlations in agents behaviour due to local
imitation processes are responsible for this
long tail. For lower values of the maximum adjustment rate
$q$, the importance of the peak with respect to
the $1/f^2$ noise is increased.

 \begin{figure}
\centerline{
\epsfxsize=80mm
\epsfig{file=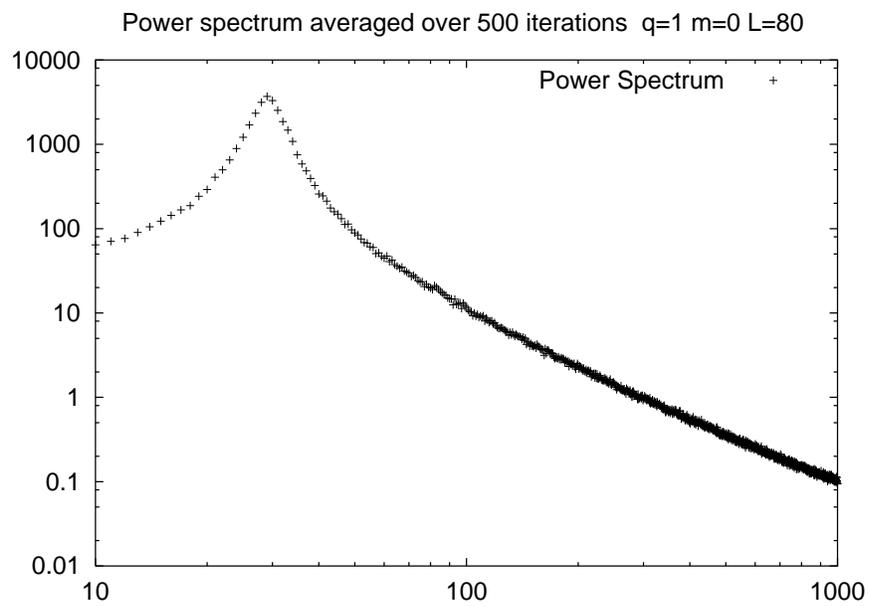,width=120mm}}
\caption{ Power spectrum in the fast adjustment regime,
for a large network ($L=80$) and fast adjustment ($q=1$)
. The frequency scale correspond to 320 updating per
 agent on average for one frequency unit}
\end{figure}

\section{Conclusions}

  The obtained results were based on very simple assumptions
on the economic network structure and on the imitation and 
adjustment process. But these results, especially the 
$1/f^2$ noise, should not depend upon the details
of these assumptions. Let us give some directions about 
the generality of our hypotheses.

 \begin{itemize}
  \item 
We based the ``voting'' process on information
processing, but this process can be also be accounted for 
on the basis of ``positive externalities''. Agents can experience
increase in the utility of equipments when their neighbours
also own such equipments.
 \item Who are the agents?
The discussion implicitly assumes 
for simplicity reasons that agents are individuals,
but the same reasoning could apply to firms 
taking decisions on purchasing goods or equipment
or even making strategic decisions. In this respect the size of the network
(number of firms) would be much smaller which could
move the dynamics towards the slow adjustment regime.
 \item The network topology:
a lattice is an extremely regular network which allows nice
pattern observation, but which cannot be considered as
a good model of a socio-economic network. In fact a lattice 
shares with real networks the property of having many
short loops, which is not the case of random nets.
Anyway the imitation model can be extended to 
structures with inhomogeneous local connectivity,
small worlds or scale free networks, 
by rewriting equation 1 using fraction of sites
with positive or negative state rather than direct
summation.
 \item
 We discussed random updating of agent states, but one can also
introduce other conditions, such as propagation
of a purchase wave as in the Weisbuch and Solomon (2000),
Weisbuch etal (2001) percolation model 
for which $1/f^2$ noise was also observed.
\end{itemize}

Let us now come to the observations.
\begin{itemize}
  \item
 The $1/f^2$ noise was expected: such fat tails
have been consistently reported in empirical data from
financial markets.  The commonly admitted reason
for the fat tails are interactions among agents.
 \item
 The periodic oscillations were unexpected, although
their origin becomes pretty evident after observation. 
 The most interesting interpretation in real life 
are business cycles. In this framework the agents
are firms and the network is the ``economy'':
the set of production, trade and services
which form the economic network.  
We here have a possible microscopic theory
of business cycles which does not suppose any external 
trigger such as innovation cycles often suggested 
by macro-economists. We probably have to take 
into account some specific features of economic
networks such as the anisotropic character
of connections (producers/users interactions are different from
competition interactions) to get more precise predictions
but some results such as the increase of the 
amplitude of activity variation with coupling 
are already within the framework of the present model. 
 
 \end{itemize}

\par
Acknowledgments: We thank Alan Kirman, Jean-Francois Laslier,
Jacques Lesourne, Jean-Pierre Nadal, Sorin Solomon,
Antonio Turiel and Jean Vannimenus for collaborations
 and helpful discussions. 
We acknowledge partial support from the FET-IST
grant of the EC IST 2001-33555 COSIN.

\bigskip

\parindent 0.5cm 

F\"ollmer H. (1974) "Random Economies with Many Interacting Agents",
 {\em Journal of Mathematical Economics}
1/1, 51-62.

Galam S., Y. Gefen and Shapir Y., (1982),
  "Sociophysics: A mean behavior model for the process of strike",
 {\em Mathem. J. of Sociology}, 9, 1-13.
 
Galam S. and Moscovici S., (1991), "Towards a theory of collective phenomena:
 consensus and attitude changes in groups",
 {\em Eur. J. of Social Psychology}, 21, 49-74.

Goldenberg J., B. Libai, S. Solomon, N. Jan, D. Stauffer, 2000,
``Marketing percolation''
Physica A, 284, 335-347.

Lesourne J. (1992), "The Economics of Order and Disorder",
Clarendon Press, Oxford

Levy H., Levy M.,  and Solomon S., (2000)
{\it Microscopic Simulation of Financial Markets},
Academic Press, New York.

Lux T. and Ausloos M., (2002),
``Market Fluctuations I: Scaling, Multi-Scaling and 
Their Possible Origins''
in A. Bunde and H.-J. Schellnhuber (Hg.): 
{\it Facets of Universality 
in Complex Systems:
Climate, Biodynamics and Stock Markets}, 
Berlin.

Orl\'ean A. (1995), "Bayesian Interactions and Collective Dynamics of
Opinions: Herd Behavior and Mimetic Contagion", {\em Journal of
Economic Behavior and Organization}, 28, 257-274. 

 Plourabou\'e F., Steyer A. and Zimmermann J.B., 
{\it Economics of
Innovation and New Technology}, {6}, 73, (1998).

Solomon S., Weisbuch G., de Arcangelis L., Jan N., and Stauffer
 D. (2000), "Social percolation models", Physica A 277, 239.

 Steyer A. and Zimmermann J.B., 
 ``Self Organised Criticality in
Economic and Social Networks: 
The case of innovation diffusion''
proceedings of the Workshop on Economics and
 Heterogeneous Interacting (2000).

Vichniac G. (1986), ``Cellular automata models of disorder and
organization'', in  {\em ``Disordered Systems and biological
organization''},
eds. Bienenstock E., Fogelman-Souli\'e, F. and Weisbuch G.,
Springer Verlag Berlin.
 
Watts D. J. and S. H. Strogatz (1998), "Collective dynamics of 'small-world' networks" {\it Nature}, 393, 440.  
 
Weisbuch G. (1990), {\em Complex Systems Dynamics}, Redwood City
(CA): Addison Wesley.

G. Weisbuch and G. Boudjema (1999), ``Dynamical aspects in the adoption
of agri-environmental measures'', Advances in Complex
Systems, 2, pp. 11-36.

G. Weisbuch and S. Solomon
"Self Organized Percolation and Critical Sales Fluctuations"
Int. Jour. Mod. Phys. C, Vol 11, No. 6, 1263-1272, (2000). 

G. Weisbuch, S. Solomon and D. Stauffer (2001)
 `` Social Percolators and Self-Organized Criticality ``
in: {\em Economics with heterogeneous interacting agents}, 
ed. by A.Kirman and J.B. Zimmermann, 
Lecture Notes in Economics and Mathematical Systems, 
Springer, Berlin- Heidelberg 2001 page 43.

\end{document}